\documentclass{Interspeech}

\usepackage[subtle]{savetrees}

\interspeechcameraready

\title{Private kNN-VC: Interpretable Anonymization of Converted Speech}

\author[affiliation={1}]{Carlos}{Franzreb}
\author[affiliation={1}]{Arnab}{Das}
\author[affiliation={1}]{Tim}{Polzehl}
\author[affiliation={2}]{Sebastian}{Möller}

\affiliation{}{German Research Center for Artificial Intelligence}{Germany}
\affiliation{}{Technical University of Berlin}{Germany}
\email{carlos.franzreb@dfki.de}
\keywords{speaker anonymization, voice conversion, interpretability}

\usepackage{comment}

\begin{document}

\maketitle

\begin{abstract}
    
Speaker anonymization seeks to conceal a speaker's identity while preserving the utility of their speech. The achieved privacy is commonly evaluated with a speaker recognition model trained on anonymized speech.
Although this represents a strong attack, it is unclear which aspects of speech are exploited to identify the speakers. Our research sets out to unveil these aspects.
It starts with kNN-VC, a powerful voice conversion model that performs poorly as an anonymization system, presumably because of prosody leakage. To test this hypothesis, we extend kNN-VC with two interpretable components that anonymize the duration and variation of phones.
These components increase privacy significantly, proving that the studied prosodic factors encode speaker identity and are exploited by the privacy attack.
Additionally, we show that changes in the target selection algorithm considerably influence the outcome of the privacy attack.
    
\end{abstract}

\section{Introduction}

Speaker anonymization has become an active field of research since the first edition of the VoicePrivacy Challenge (VPC) \cite{tomashenko_introducing_2020}, which provides an evaluation protocol to assess the strength of anonymization systems.
The evaluation simulates an attack on the anonymized speech, leveraging a speaker recognition model trained to discriminate anonymized speakers by learning their characteristics. In the VPC task, the spoken content is ignored. Identifying speakers from text is a different task with its own challenges \cite{sinha_safeguarding_2024}.
Privacy is measured in terms of its trade-off with utility, which should be preserved throughout anonymization. How utility is defined depends on the use case; the VPC 2024 \cite{tomashenko2024vpc} focuses on human perception, measuring utility with speech and emotion recognition models trained on original speech.

Voice Conversion (VC) models change the speaker identity of speech from a source speaker to a target speaker \cite{sisman_overview_2021}. Their goal is to produce speech with great target similarity, which is complementary to the anonymization's goal of concealing the source speaker. kNN-VC \cite{baas23_interspeech} is a popular VC model because of its simplicity, versatility and the high quality of the converted speech. It works for multiple languages without re-training the vocoder \cite{franzreb_towards_2024}, and also for any target speaker for which sufficient speech data is available (5 minutes according to their experiments).

Despite achieving state-of-the-art results on subjective target speaker similarity, kNN-VC fails to conceal the source speaker's identity according to several publications that use the VPC 2024 objective evaluation \cite{franzreb_towards_2024, cai2024privacy, ghosh_anonymising_2024, das2024comparing}. Most authors argue that the identity leakage arises from prosodic features, i.e. the intonational and rhythmic aspects of speech \cite{jurafsky_speech_2000}. 
Recent work has shown that prosody is leveraged by the attacker of the VPC evaluation to identify speakers by experimenting with deep prosodic representations \cite{gengembre_disentangling_2024}, pitch contours \cite{champion_anonymizing_2023} and phone durations \cite{jin_speaker_2009, tomashenko_analysis_2025}, making prosody a good candidate for explaining kNN-VC's poor privacy.
However, this hypothesis has not been empirically proven yet.
More generally, identifying the aspects of speech which are exploited by the attacker to identify anonymized speakers is an active field of research.
This is due to the complex nature of speech, which makes model explainability challenging \cite{fucci2024explainability}, as well as to the deep architectures used for anonymization, whose behavior is unpredictable and hard to understand \cite{rudin_stop_2019}.

Anonymizing different aspects of speech independently can help us understand how the attacker identifies speakers, as we can perform ablation studies looking at each aspect individually, as well as studying their possible combinations.
\cite{meyer_prosody_2023} proposes such an anonymization system, conditioning the vocoder on the phone durations as well as on the pitch and energy contours of the source speech. Pitch and energy contours are anonymized by applying a random percentage offset to each individual value.
However, the system achieves perfect privacy on the VPC 2022 evaluation even when using the original prosodic features, leaving unclear how the anonymization of pitch and energy impacts privacy.

We propose a similar approach to enhance the anonymization capabilities of kNN-VC, targeting two aspects of prosody: the duration and variation of phones.
We anonymize phone durations by predicting new durations from the phonetic transcript of the source speech, following an approach used for text-to-speech \cite{ren2021fastspeech}.
To manipulate phonetic variation, we quantize the target speech with k-means clustering. Fewer clusters result in a lower resolution of information available for the neighbor selection of kNN-VC, thereby restricting individual expressive characteristics.
The two aspects can be anonymized independently, and in different levels of strength, allowing us to assess their effect in the results of the VPC 2024 evaluation.

In our experiments, we observed that target selection significantly impacted the privacy outcomes in the VPC 2024 evaluation, even though our anonymization system performs consistently across all target speakers, as confirmed by the utility evaluation. This discrepancy underscores the sensitivity and nuance of privacy assessment. We conduct additional experiments to analyze the attacker's behavior in relation to target speaker selection and discuss the broader implications of our findings.

The code and weights for the proposed anonymization system are open-source \footnote{\url{https://github.com/carlosfranzreb/private_knnvc}}. It is implemented in SpAnE \cite{franzreb_comprehensive_2023}, where the inference takes place, and evaluated with the VPC 2024 \cite{tomashenko2024vpc}.

\section{Anonymization system}

\begin{figure}
    \centering
    \includegraphics[width=\linewidth]{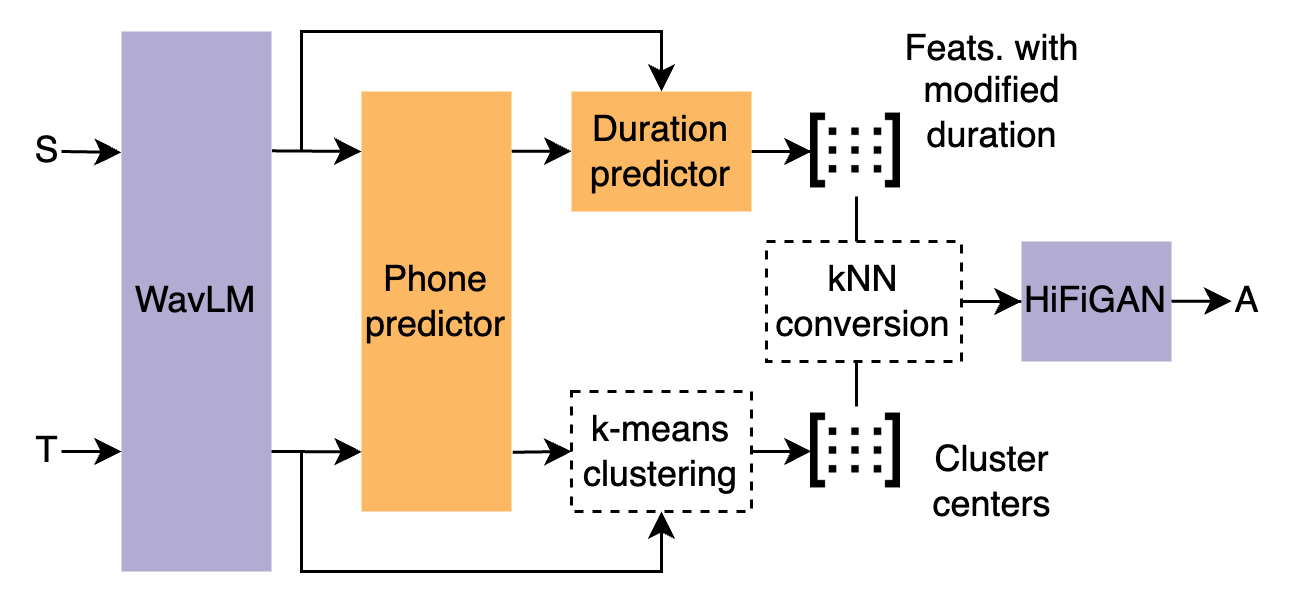}
    \caption{Private kNN-VC: purple components are the same as those of kNN-VC. $S$ is the source speech, $T$ is the target speaker's speech, and $A$ is the anonymized speech.}
    \label{fig:model}
\end{figure}

kNN-VC \cite{baas23_interspeech} is an any-to-any VC model. It first extracts features from both source and target speech using WavLM \cite{wavlm_ieee}, a speech language model. Then, each source feature is replaced with the average of its four nearest target features according to cosine similarity.
This works because the representation space of speech language models like WavLM is heavily influenced by phones \cite{sicherman_analysing_2023}.
The converted features are synthesized with a HiFi-GAN vocoder \cite{kong_hifi-gan_2020} trained on data that simulates the kNN conversion by replacing the features of each training sample with others from the same speaker.

We enhance kNN-VC for speaker anonymization by adding two trained predictors (see Figure \ref{fig:model}). The phone predictor assigns a phone to each WavLM feature, while the duration predictor estimates the duration of each phone in a phonetic transcript. The transcript is obtained by removing consecutive occurrences from the sequence of predicted phones.
Both predictors share the same convolutional architecture as the variance predictors from FastSpeech2 \cite{ren2021fastspeech}.

\subsection{Phone predictor}

The phone predictor is trained with the same data used by SUPERB \cite{yang_superb_2021} for their phone recognition task. It consists of the Librispeech \textit{train-clean-100} dataset \cite{panayotov_librispeech_2015}, for which the aligned phones were predicted with a g2p model. The alphabet is based on the CMU dictionary\footnote{\url{http://www.speech.cs.cmu.edu/cgi-bin/cmudict}}: there are 41 phones and lexical stress is ignored. 80\% of the data is used for training, 10\% for validation and 10\% for testing.

We train this predictor with two losses: cross entropy and connectionist temporal classification (CTC) \cite{graves2006connectionist}. Cross entropy considers the accuracy of each prediction, while CTC considers the whole phonetic transcript, without duplicates. Combining these two losses, the resulting phone predictor is both accurate and produces a clean phonetic transcript. (ACC=92\% and PER=2\% on the held-out test set). It took 4 hours to train on one NVIDIA RTXA6000 GPU.

\subsection{Controlling the phonetic variation}

Recent work on voice conversion shows that quantizing features of speech language models can be used for manipulating voice characteristics \cite{zhou_zero-shot_2024} and disentangling prosodic features \cite{sim_skqvc_2024}.
Inspired by their approaches, we use the phone predictor to quantize the target features. We group the target features based on their predicted phones, and then cluster each group with k-means \cite{hartigan1979k}. The cluster centers are used for the conversion step instead of the target features.
The number of clusters determines how many representations of each phone are available for the conversion, which ultimately controls the phonetic variation of the anonymized speech.
As the clustering already aggregates features, each source feature is exchanged with its nearest cluster center, rather than averaging the four nearest neighbors as in kNN-VC.

\subsection{Duration predictor}

The duration predictor is trained like the one from FastSpeech2 \cite{ren2021fastspeech}, with 24 hours of speech from a single LJSpeech speaker \cite{ljspeech17}.
and the mean squared error (MSE) loss. For a given speech sample, the true duration of each phone is its number of consecutive predictions by our phone predictor. Improvement on the validation set stagnates after 20 epochs. The test loss is MSE=4.4. Once the true durations have been computed, training the duration predictor takes less than a minute in an Apple M1 chip.
At inference time, the predicted durations are applied by evenly sampling features from the actual range of source features. For example, if a phone has a true duration of 5 features, but the predicted duration is 3, we remove the second and fourth features. If the predicted duration is larger than the true duration, we duplicate the intermediate features.

Since the duration predictor is trained on data from a single speaker, its predictions are not conditioned on the target speaker selected for anonymization, hindering target fidelity.
However, our goals are to protect the source speaker's identity and to preserve the speech's utility; neither is compromised by poor target fidelity.

\subsection{Nomenclature for model configurations}

The degrees of manipulation for the duration and variation of phones are defined by two parameters.
The first parameter is the weight $w \in [0,1]$, assigned to the predicted durations \vec{p}. They are summed with the true durations \vec{t}, whose weight is $1-w$, to produce the durations \vec{d}, as shown in Equation 1 below.
The second parameter is the number of clusters used to quantize the target features of each phone.
The two parameter values are used to identify each configuration. For clarity, $w$ is expressed as a number between 1 and 10. For example, a configuration with $w=0.7$ and 8 clusters is named $(7$-$8)$.

\begin{equation}
    \vec{d} = w \cdot \vec{p} + (1-w) \cdot \vec{t}
\end{equation}

\section{Evaluation}

We evaluate our anonymization system with the official implementation of the VPC 2024 \cite{tomashenko2024vpc}.
As target speakers, we use 100 speakers from the LibriTTS \cite{zen19_libritts} \textit{train-other-500} dataset, evenly distributed across genders.
The target selection algorithm randomly selects a target speaker while preserving the source speaker's gender. It is applied independently to each utterance, allowing different utterances from the same source to be anonymized with different target speakers.

Privacy is measured with a speaker recognition model trained on anonymized Librispeech \textit{train-clean-360} data \cite{panayotov_librispeech_2015}, and evaluated on a verification setup with the Librispeech \textit{test-clean} dataset. Speaker embeddings of trial and enrollment utterances are compared with cosine similarity.
The evaluation metric is the equal error rate (EER). EER=0\% means that all pairs of samples were correctly recognized, and EER=50\% is the best achievable result in terms of privacy.
We average results for male and female trials, which are reported separately by the VPC 2024 framework. For a few configurations, the male EER was up to 6\% higher than the female EER; otherwise, the results were similar. 
For the averaging to be fair, EER values above 50\% are replaced by their subtraction from 100 (i.e. $100-$EER). EERs exceeding 50\% only occur when the EER is close to 50\%, because of the thresholding procedure.

The VPC 2024 measures utility with models trained on original data. Intelligibility is estimated with the word error rate (WER) of a speech recognition model trained on LibriSpeech. Emotion preservation is estimated with the unweighted average recall (UAR) of an emotion recognition model trained and evaluated on the speech samples of IEMOCAP \cite{busso2008iemocap}.
Recent experiments have revealed a significant discrepancy between objective and subjective utility measures in anonymized speech \cite{vpc22_results}. Note that in our analysis, the terms intelligibility and emotion preservation are used exclusively to describe the outcomes of the objective evaluation, without implying a direct correlation to subjective perception.

\section{Experiments}

\begin{figure*}[th]
\minipage{0.34\textwidth}
  \includegraphics[width=\linewidth]{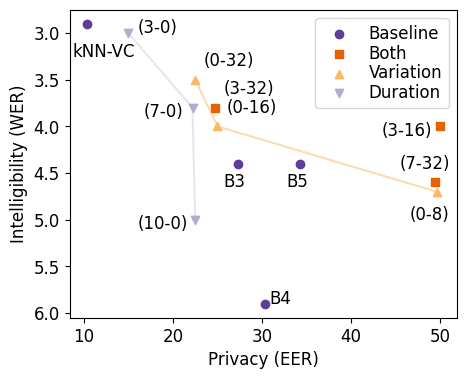}
  \caption{Privacy vs. intelligibility}
  \label{fig:eer_vs_wer}
\endminipage
\hfill
\minipage{0.34\textwidth}
  \includegraphics[width=\linewidth]{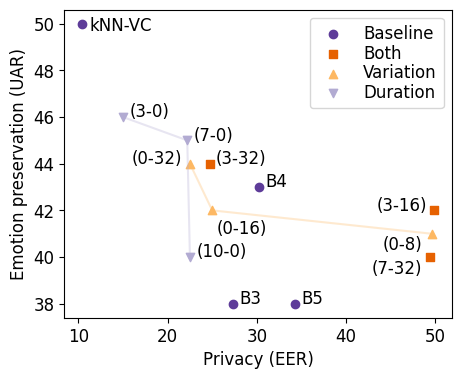}
  \caption{Privacy vs. emotion preservation}
  \label{fig:eer_vs_uar}
\endminipage
\hfill
\minipage{0.3\textwidth}
  \includegraphics[width=\linewidth]{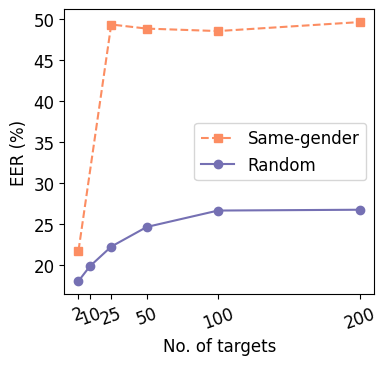}
  \caption{No. of targets vs. privacy}
  \label{fig:number_targets}
\endminipage
\hfill
\end{figure*}

We experiment with different configurations of our model to investigate how each of the prosodic aspects (duration and variation of phones) affect the privacy attack. For each parameter, we have selected four values.
Phonetic variation is anonymized with 32, 16 or 8 clusters, or not anonymized, which we define as zero in our configuration identifiers. Zero means that no clustering is performed, i.e. all available target features are candidates for the conversion. Using fewer than eight clusters is possible, but it adversely affects intelligibility.
For the duration weight $w$, we use values 0, 0.3, 0.7 and 1. Zero means that the phone durations are not modified, and one means that the predicted phone durations are used without summing them with the true durations at all.
Each experiment takes roughly 5 hours on a RTXA6000 GPU.
We compare the proposed system with the original kNN-VC and the baselines from the VPC 2024 \cite{tomashenko2024vpc}.

\subsection{Phone duration and variation}

Figures \ref{fig:eer_vs_wer} and \ref{fig:eer_vs_uar} depict the privacy-utility trade-offs computed with the VPC 2024 framework. Figure \ref{fig:eer_vs_wer} shows a scatter plot of privacy against intelligibility. The ideal anonymization system would output highly intelligible speech while enhancing privacy. In the plot, such a system would lie in the top-right corner. Figure \ref{fig:eer_vs_uar} shows privacy against emotion preservation. In this plot, the ideal anonymization system would also lie in the top-right corner.

A slight anonymization of the phone duration ($w=3$) improves the privacy of kNN-VC from 10\% to 15\% in terms of EER.
Increasing the weight further increases the EER up to 22.5\%. The difference in EER between $(7$-$0)$ and $(10$-$0)$ is negligible. These results show that phone durations encode speaker identity, and that the attacker exploits them to recognize the source speakers.
Both intelligibility and emotion preservation decrease monotonically for increasing values of $w$.
Compared with the VPC baselines, $(7$-$0)$ ranks second in terms of intelligibility with WER=4\%, and third in emotion preservation with UAR=43\%, while providing better privacy than the baselines with better utility (22.2\% EER, against 6.1\% of B1 and 4.5\% of B2).
We attribute the loss in intelligibility to the errors made by our predictors; emotion preservation decreases because our duration predictor does not consider the emotion of the source speech, which influences rhythm and stress patterns.

Reducing the phonetic variation of the anonymized speech is more effective at protecting the source speakers' identities than altering phone durations. $(0$-$32)$, which constitutes the smallest constraint possible of the phonetic variation in our experiments, achieves the same EER as the complete anonymization of phone durations (EER=22.5\%). Reducing the number of clusters further continually improves the EER up to 49.6\% for 8 clusters.
Regarding utility, reducing the phonetic variation has a lower impact on intelligibility than altering the phone durations, with a max. WER of 4.7\%. The degradation of emotion preservation is comparable for both anonymization techniques.
Our experiments where both the variation and duration of phones are anonymized show that these two techniques are complementary. $(3$-$16)$ achieves an EER of 49.9\% and provides the same utility as $(0$-$16)$ (WER=4\%, UAR=42\%). Variation can be increased while increasing phone duration anonymization, without a loss in privacy, as shown by $(7$-$32)$.

\subsection{Best configurations}

\begin{table}[th]
  \caption{VPC 2024 evaluation results. A suffix in our configuration identifiers means that a different target selection algorithm was used: $c$ stands for \textit{cross-gender}; $r$ stands for \textit{random}, and $d$ for \textit{disjoint}.}
  \label{tab:vpc2024}
  \centering
  \begin{tabular}{ *4c }
    \toprule
        \textbf{Model} &
        \textbf{Emotion} $\uparrow$ &
        \textbf{Intelligibility} $\downarrow$ &
        \textbf{Privacy} $\uparrow$ \\
         & UAR (\%) & WER (\%) & EER (\%) \\
    \midrule
        Original & 71  & 1.8 & 4.6 \\
    \midrule
        B1 & 43 & 2.9 & 6.1 \\
        B2 & 53 & 10.0 & 4.5 \\
        B3 & 38 & 4.4 & 27.3 \\
        B4 & 43 & 5.9 & 30.3 \\
        B5 & 38 & 4.4 & 34.3 \\
        B6 & 36 & 9.1 & 21.1 \\
        kNN-VC & 50 & 2.9 & 10.4  \\
    \midrule
        Ours $(0$-$32)$ & 44 & 3.5 & 22.5 \\
        Ours $(0$-$8)$ & 41 & 4.7 & 49.6 \\
        Ours $(3$-$16)$ & 42 & 4.0 & 49.9 \\
        Ours $(7$-$8)$ & 39 & 5.7 & 48.5 \\
    \midrule
        Ours $(0$-$32)_c$ & 42 & 3.8 & 23.4 \\
        Ours $(0$-$8)_c$ & 38 & 4.7 & 49.3 \\
        Ours $(3$-$16)_c$ & 40 & 4.2 & 49.0 \\
        Ours $(7$-$8)_c$ & 37 & 5.9 & 48.8 \\
    \midrule
        Ours $(0$-$32)_r$ & 42 & 3.7 & 17.8 \\
        Ours $(0$-$8)_r$ & 39 & 4.8 & 18.3 \\
        Ours $(3$-$16)_r$ & 40 & 4.3 & 19.1 \\
        Ours $(7$-$8)_r$ & 37 & 5.7 & 26.6 \\
    \midrule
        Ours $(10$-$8)_r$ & 35 & 7.3 & 29.4 \\
    \midrule
        Ours $(0$-$8)_{d,1}$ & 40 & 4.7 & 48.4 \\
        Ours $(0$-$8)_{d,2}$ & 40 & 4.8 & 49.4 \\
    \bottomrule
  \end{tabular}
\end{table}

The configurable nature of our model makes it adaptable to different privacy requirements. The VPC also evaluates models in this way, to account for different use cases of anonymization \cite{tomashenko2024vpc}. The motivation behind this is the trade-off between privacy and utility. Use cases that don't require as much privacy may therefore benefit from better utility.
Table \ref{tab:vpc2024} shows the VPC 2024 evaluation results for the baselines and some of our configurations. kNN-VC offers the best utility, but little privacy. Our extensions of it offer gradual improvements in privacy. Privacy can be improved by reducing the phonetic variation, with $(0$-$32)$ doubling the EER of kNN-VC, and $(0$-$8)$ achieving almost perfect privacy according to the results of this evaluation. Removing phone duration anonymization makes the model simpler and faster, as the duration predictor is not needed.
However, for use cases that require more phonetic variation, the addition of the phone duration anonymization allows us to increase the number of clusters without a loss in privacy, as shown by $(3$-$16)$.

\section{Effect of target selection on privacy}

Our model picks target speakers randomly, but preserves the gender of the source speaker. Randomness ensures \textit{unlinkability}, i.e. that source speakers cannot be identified by attacking solely the target selection algorithm \cite{champion_anonymizing_2023}. Preserving gender is a reasonable requirement for many use cases, as it facilitates fairer evaluations of the anonymized datasets, helping to identify and mitigate potential biases.
Previous work also reports improved target speaker similarity in other extensions of kNN-VC when same-gender conversion is enforced \cite{liu_two-stage_2024}.
We observe similar results: the configuration identifiers in Table \ref{tab:vpc2024} that have $c$ as a suffix refer to experiments where cross-gender conversion is enforced, instead of same-gender conversion. Cross-gender conversion decreases emotion preservation between 2\% and 3\% UAR. Intelligibility is also consistently worse, but the difference is marginal. Privacy is not affected significantly by the gender of the target speaker.

We have also run the same experiments as before, but with completely random target selection, ignoring gender. The results are shown in Table \ref{tab:vpc2024}, where the configuration identifiers have $r$ as a suffix. The EER of all configurations of our model which achieved EERs close to 50\% under the same-gender constraint decrease by over 20\%. The highest EER achieved for completely random target selection is 29.4\%, obtained with $(10$-$8)_r$, the strongest configuration possible in terms of privacy in our experiments. 
In this section, we perform two additional experiments to analyze how target selection affects the efficacy of the attack.

\subsection{Effect of gender constraint on privacy}

As the privacy achieved by same-gender and cross-gender conversion is similar, we hypothesize that the attacker is being confused by the two disjoint sets of targets, regardless of the gender distribution of the two sets.
We test this by splitting the 100 target speakers into two groups, where each group has 25 female and 25 male members. One group is used to anonymize female source speakers, and the other for male source speakers. In this way, there are still two disjoint sets of targets, but they both comprise an even distribution across genders.
We have run this experiment for the two possible combinations of target set and source gender for the configuration $(0$-$8)$. The results of this experiment are reported in Table \ref{tab:vpc2024}, with the suffixes $d,1$ and $d,2$.
These experiments yield the same privacy results as when a gender constraint is placed on the target selection: the EERs are close to 50\%. Thus, it is not gender that is confounding the attacker, but having two disjoint sets of target speakers, each of them being used to anonymize roughly half of the training and evaluation samples.

\subsection{Effect of number of targets on privacy}

The results presented until now suggest that the attacker is learning particularities about each target, instead of looking for remaining information about the source speakers in the anonymized speech. Its ability to recognize speakers is therefore likely to be impacted by the number of targets used. To test this effect, we have anonymized our model $(7$-$8)$ with different numbers of targets. The privacy results are shown in Figure \ref{fig:number_targets}. When target selection is completely random, privacy saturates at an EER of 27\%, achieved with 100 targets. When the same-gender constraint is added to the selection algorithm, privacy already nears 50\% for 25 target speakers.

\section{Limitations}

We identify three areas of improvement for the proposed system: privacy, interpretability and emotion preservation.

The large difference in EER when the target selection algorithm is changed leads us to believe that there is still personally identifiable information (PII) in the anonymized speech produced by our system, although the attacker fails to exploit it for certain target selections. Related work has shown that some selection algorithms may confound the attacker, leading to an overestimation of privacy \cite{champion_anonymizing_2023}. Same-gender selection seems to be such an algorithm in our case, but only for certain configurations.
Our current experiments can only discard phone duration as the cause of the privacy leakage; timbre and phonetic variation could still contain PII.
It is also unclear whether certain target speakers provide better anonymization than others. Acoustic differences between source and target speakers could impact the results of the kNN conversion.

Differences between target speakers also hinder the interpretability of phonetic variation. We cannot ensure that the clustering algorithm constrains phonetic variation in the same way for all target speakers. The effect of these differences in the results of our experiments is unclear.
Interpretability could be further improved by increasing the granularity of phonetic variation, which can be replaced by pitch, energy and other prosodic aspects.

Lastly, we do not consider the emotion of the source speech in our anonymization system. Addressing this should improve the utility of the anonymized speech.
To enhance emotion preservation without leaking speaker identity, we need to recognize emotions, and to model how the target speakers expresses them. For example, the duration predictor could be trained with an emotional dataset, conditioning the durations on the emotion of each sample.

\section{Conclusion}

We propose an interpretable extension of kNN-VC for speaker anonymization.
Each anonymization technique is specific and configurable, allowing us to research its effect on utility and privacy.
Our experiments show that the variation and duration of phones encode speaker identity, which the VPC 2024 attacker exploits to identify speakers. 
Constraining the phonetic variation provides stronger anonymization than altering the phone durations. Their combination allows our system to cater to different utility requirements without losing privacy.
Unexpectedly, our experiments demonstrate that target selection significantly influences objective privacy evaluations, indicating that residual PII persists in the anonymized speech.

\section{Acknowledgements}

Funded by Federal Ministry of Education and Research, Germany (BMBF 16KIS2048).

\bibliographystyle{IEEEtran}
\bibliography{references, other_references}

\end{document}